\documentclass[11pt]{article}
\usepackage{amssymb, amsmath,amsthm}

\DeclareMathOperator{\Id}{Id}

\newtheorem{theorem}{Theorem}[section]
\newtheorem{thm}[theorem]{Theorem}

\newtheorem{lem}[theorem]{Lemma}
\newtheorem{prop}[theorem]{Proposition}

\begin{document}

\title{EXPLICIT SOLUTIONS OF LAPLACE EQUATIONS WITH INVERSE SQUARE AND SINGULAR P\"OSCHL-TELLER POTENTIALS}
\author{ Mohamed Vall OULD MOUSTAPHA}
\maketitle

\begin{abstract} In this article we give in analytical closed form the solutions of the Direchlet problems for the Laplace equations with inverse square and singular P\"oschl-Teller potentials.
\end{abstract}

\begin{center}
\section{Introduction }
\end{center}
The Laplace equation is used in many contexts. For example, 
 in potential theory, in electrostatics and in complex analysis as well as in
 Riemannian geometry.\\ 
This study shows in the unified way the explicit formulas for the following three equations of Laplace type:\\
{\bf a)} The Laplace equation associated to the Schr\"odinger operator with the inverse square potential on $R^+$
 $$  \left \{
    \begin{aligned}
      L_{\nu}u(Y, X)+\partial_Y^2 u(Y, X)=0,
      &\qquad (Y, X)\in \mathbb{R_+}\times {\mathbb{R_+}}
      \\
      u(0,X)=u_0(X)\in
      C_0^\infty({\mathbb{R}_+^\ast}).
    \end{aligned}
  \right.\eqno(1.1)$$
where
$$ L_\nu =\frac{\partial^{2}}{\partial X^{2}}+ \frac{1/4-\nu^2}{X^2},
\eqno(1.2) $$
is the Schr\"odinger operators with inverse square potential.

{\bf b) } The Laplace type equations associated to the Schr\"odinger operators with the singular trigonometric P\"oschl-Teller  potentials
$$
  \left \{
    \begin{aligned}
      L^T_{\nu}v(y, \theta)+\partial_y^2 v(y, \theta)=0,
      &\quad(y, \theta)\in \mathbb{R}_+\times [0,\pi]
      \\
      v(0,\theta)=v_0(\theta)\in
      C_0^\infty([0,\pi]).
    \end{aligned}
  \right.\eqno(1.3) $$
where
$$
  L^T_{\nu}=\frac{\partial^{2}}{\partial \theta^{2}}+ \frac{1/4-\nu^2}{16 \sin^2 \theta},
  \eqno(1.4) $$
is the Schr\"odinger operators with singulars trigonometric P\"oschl-Teller potential.

{\bf c) }The Laplace type equations associated to the Schr\"odinger operators with the singular hyperbolic P\"oschl-Teller  potentials

$$
  \left \{
    \begin{aligned}
      L^H_{\nu}w(y, x)+\partial_y^2 w(y, x)=0,
      &\quad(y , x)\in \mathbb{R_+}\times {\mathbb{R_+}}
      \\
      w(0,x)=w_0(x)\in
      C_0^\infty({\mathbb{R}_+}).
    \end{aligned}
  \right.
\eqno(1.5) $$
where
$$
  L^H_{\nu}=\frac{\partial^{2}}{\partial x^{2}}+ \frac{1/4-\nu^2}{16 \sinh^2 x},
 \eqno(1.6) $$

is the Schr\"odinger operators with singulars hyperbolic P\"oschl-Teller potential.
That is the researcher computes explicitly the Schwartz  integral kernels of the following Poisson semi-groups
$\exp{\left(-Y\sqrt{- L_\nu}\right)}$, $\exp{\left(-\theta\sqrt{- L_\nu^T}\right)}$ and  $\exp{\left(-y\sqrt{-L_\nu^H}\right)}$.\\ The  Schr\"odinger operator with inverse square potential $L_\nu$
arises in the contexts of the Schr\"odinger equation in non relativistic quantum mechanics $[15
]$, the molecular physics and the quantum cosmology as well as the linearization of combustion models [1]and [2].\\
For example, the Hamiltonian for a spinzero particle in Coulomb field gives rise to a Schr\"odinger operator involving the inverse square potential $[5]$.\\ The Schr\"odinger equation with the P\"oschl-Teller potentials was studied a long time ago. First in 1882 by Darboux [8], then independently by  P\"oschl and Teller in 1933 [13].\\
The importance of the P\"oschl-Teller potentials in mathematics, physics and chemistry is well known. These potentials
represent one of the most studied anharmonic systems. On the one hand, the Schr\"odinger equation with this potential plays an important role in many body integrable systems [6] and [11] in soluton mathematics, from which
the multi-soluton solutions of the nonlinear Korteweg-de Vries (KdV) equation can be explicitly constructed [9] and [16], and in the Hartree mean field equation of many body systems interacting  trought a delta force [7
] and [17]. On the other hand the Schr\"odinger equation with singular trigonometric and hyperbolic  P\"oschl-Teller potentials
can be also regarded as Schr\"odinger equations with inverse square potential on the one dimensional  spherical and the hyperbolic spaces respectively. \\
The rest of the paper is organised as follows, this section is ended by recall about the Hankel transform, in section 2, the solution of Laplace equation with inverse square potential (1.1) is given  in a closed form in terms of 
Legendre functions of the second kind. In sections 3 and 4  the solutions of Laplace equations with singular trigonometric and hyperbolic P\"oschl-Teller potentials are given in a closed form.\\
Now some facts about the Hankel transform are given (see $[4],[12]$).\\
When $\nu>-1$ , the Hankel transform of order $\nu$  is defined as
$$
\left({\cal H}_\nu f\right)(\Omega)=\int_0^\infty
(X\Omega)^{1/2}J_\nu(X\Omega)f(X)dX
\eqno(1.7) $$
where $f\in C_0^\infty\left({R_+^\ast}\right)$ and $J_\nu$
is the Bessel function of the first kind and order~$\nu$.

\begin{prop}(see$[4]$ and $[12]$) The Hankel transform has the following properties:
  \begin{enumerate}
  \item[i)] ${\cal H}_\nu^2=\Id$.
  \item[ii)] ${\cal H}_\nu $ is self-adjoint.
  \item[ii)] ${\cal H}_\nu $ is an $L^2$-isometry
  \item[iii)]${\cal H}_\nu L_\nu^E=-\Omega^2{\cal H}_\nu$.
  \end{enumerate}
\end{prop}
\section{Laplace equation with inverse square potential on $R^+$}
\begin{thm} The Direchlet problem  $(1.1)$ for the Laplace equation with inverse square potential on the Euclidian line has the unique solution given by

$$
u(Y,X)=\int_0^\infty {\cal P}_\nu(Y, X ,X')u_0(X')dX'\eqno(2.1) $$
where
$$
 {\cal P}_\nu(Y, X ,X')=\frac{-4 Y}{\sqrt{Y^2+(X+X')^2}\sqrt{Y^2+(X-X')^2}}Q_{\nu-1/2}^1\left(\frac{Y^2+X^2+X'^2}{2X X'}\right)\eqno(2.2) $$
where  $Q_{\nu-1/2}^ \mu$ is the associated Legendre function of the second kind given in terms of the Gauss hypergeometric as\\
$
  Q_{\nu-1/2}^ \mu(z)=\frac{\sqrt{\pi}e^{i\mu\pi}\Gamma(\mu+\nu+1/2)}{2^{\nu+1/2}\Gamma(\nu+1)} z^{-\mu-\nu-1/2}(z^2-1)^{\mu/2}\times$
$$F\left(\frac{\mu+\nu+1/2}{2}, \frac{\mu+\nu+3/2}{2},
  \nu+1; \frac{1}{z^2}\right)
\eqno(2.3) $$
The Gauss hypergeometric function is defined by:
$$
  F(a,b,c;z)=\sum_{n=0}^{\infty}\frac{(a)_n(b)_n}{(c)_n n!}z^n,
  \quad |z|<1,
\eqno(2.4) $$
where as usual $(a)_n$ is the Pochhamer symbol defined by
$$
  (a)_n=\frac{\Gamma(a+n)}{\Gamma(a)}
\eqno(2.5) $$
and $\Gamma$ is the classical Euler function.
\end{thm}
{\bf Proof} By the Hankel transform I can transform the Direchlet problem $(1.1)$ into the following

$$
  \label{1}
  \left \{
    \begin{aligned}
      -\omega^2({\cal H}u)(Y, \omega)+\partial_Y^2 ({\cal H}u)(Y, \omega)=0,
      &\qquad (Y, X)\in \mathbb{R_+}\times {\mathbb{R_+}}
      \\
      ({\cal H}u)(0, \omega)=({\cal H}u_0)(\omega)\in
      C_0^\infty({\mathbb{R}_+^\ast}).
    \end{aligned}
  \right.\eqno(2.6)
$$
The solution of $(2.6)$ is given by $$ ({\cal H}u)(Y, \omega)=\exp{(- Y\omega)}({\cal H}u_0)(\omega)\eqno(2.7)$$
Using the inverse Hankel transform and
in view of the following asymptotic formulas for the Bessel functions ([10] p.134):
$$J_\nu(x)\approx \frac{x^\nu}{2^\nu\Gamma(1+\nu)}; x\rightarrow 0\eqno(2.8)$$
$$J_\nu(x)\approx \sqrt{\frac{2}{\pi x}}\cos(x-(\nu\pi/2)-(\pi/4)); x\rightarrow \infty \eqno(2.9)$$
we can use the Fubini theorem to obtain
$$ u(Y,X)=\int_0^\infty u_0(X')\int_0^\infty (X X')^{1/2}J_\nu(X \Omega) J_\nu(X' \Omega) e^{-Y \Omega} \Omega d\Omega dX'\eqno(2.10)$$
Now using  the formula in $([14], p.286-287)$
$$\int_0^\infty e^{-p x} J_\nu(a x) J_\nu(b x) x d x=\frac{-p k^2(ab)^{-3/2}}{\sqrt{1-k^2}}Q_{\nu-1/2}^1\left(\frac{2-k^2}{k^2}\right)\eqno(2.11)
$$
${\cal R} e (\nu)>-1/2$, ${\cal R} e p>|{\cal I}m a|+|{\cal I} m b|$, and
$k=2\sqrt{a b}\left(p^2+(a+b)^2\right)^{-1/2}$, we obtain the result of the theorem.\\
We end this section by the following lemma
\begin{lem} Set $X=\varphi(\xi)+\psi(\eta)$, $Y=\varphi(\xi)-\psi(\eta)$ then we have:\\
i) For $F\in C^2 $ and $\varphi, \psi\in C^1$ the following formula holds
$$\left[\frac{\partial^2 F}{\partial X^2}+\frac{\partial^2 F}{\partial Y^2}+V(X)\right]F(X,Y)=\left(\varphi'(\xi)\psi'(\eta)\right)^{-1}\left[4\frac{\partial^2 }{\partial \xi\partial\eta}+\varphi'(\xi)\psi'(\eta)V\left(\varphi(\xi)+\psi(\eta)
\right)\right]G(\xi,\eta)$$

ii) For $\varphi(\xi)=\tan\xi$ and $\psi(\eta)=\tan\eta $ with $\xi=\frac{\theta+iy}{2}$ and $\eta=\frac{\theta-iy}{2}$, the 
following formula hold:
$$X_1=:\varphi(\xi)+\psi(\eta)=\tan\xi+\tan\eta=\frac{2\sin\theta}{\cos\theta+\cosh y}$$
$$Y_1=:\varphi(\xi)-\psi(\eta)=\tan\xi-\tan\eta=\frac{2i\sinh y}{\cos\theta+\cosh y}$$

iii) For $\varphi(\xi)=\tanh\xi$ and $\psi(\eta)=\tanh\eta $ with $\xi=\frac{x+y}{2}$ and $\eta=\frac{x-y}{2}$  the following formulas hold:
$$X_2=:\varphi(\xi)+\psi(\eta)=\tanh\xi+\tanh\eta=\frac{2\sinh x}{\cosh x+\cosh y}$$
$$Y_2=:\varphi(\xi)-\psi(\eta)=\tanh\xi-\tanh\eta=\frac{2\sinh y}{\cosh x+\cosh y}$$
\end{lem}
The proof of this lemma is simple and hence is left to the reader.
\section{Laplace equation with singular trigonometric P\"oschl-Teller potential}
\begin{thm} The Direchlet problem $ (1.3)$ for Laplace equation with singular trigonometric P\"oschl-Teller on the Spherical line has the unique solution given by
$$
v(y,\theta)=\int_0^\pi {\cal P}_\nu^T(y, \theta ,\theta')v_0(\theta')d\theta'
\eqno(3.1)$$
where
$$
  {\cal P}_\nu^T(y, \theta, \theta')= {\cal P}_\nu(Y_1, X_1, \theta')\eqno(3.2)
$$
where ${\cal P}_\nu$  is the Schwartz integral kernel of the Poisson semi-group $\exp{\left(-Y\sqrt{-L_\nu}\right)}$ with inverse square potential given in $(2.2)$
with  $Y_1$ and $X_1$ as in Lemma 2.2

\end{thm}

{\bf Proof}
We take
 $ V(X_1,  Y_1)=\frac{1/4-\nu^2}{X_1^2}$  and set
$$X_1=:\varphi(\xi)+\psi(\eta)=\tan\xi+\tan\eta; Y_1=:\varphi(\xi)-\psi(\eta)=\tan\xi-\tan\eta\eqno(3.3)$$
with $\xi=\frac{\theta+iy}{2}$ and $\eta=\frac{\theta-iy}{2}$ in Lemma 2.2 we obtain
$$\left[\frac{\partial^2 F}{\partial X_1^2}+\frac{\partial^2 F}{\partial Y_1^2}+\frac{1/4-\nu^2}{X_1^2}\right]F=\left((1+\tan^2\xi)(1+\tan^2 \eta)\right)^{-1}\left[4\frac{\partial^2 }{\partial \xi\partial\eta}+\frac{1/4-\nu^2}{\sin^2(\xi+\eta)}
\right]F\eqno(3.4)$$
we obtain
$$\left[\frac{\partial^2 }{\partial X_1^2}+\frac{\partial^2 }{\partial Y_1^2}+\frac{1/4-\nu^2}{X_1^2}\right]F=16 \cos^2((\theta+iy)/2)\cos^2((\theta-iy)/2)\left[\frac{\partial^2 }{\partial \theta^2}+\frac{\partial^2 }{\partial y^2}+\frac{1/4-\nu^2}{16 \sin^2\theta}\right]F\eqno(3.5)$$
To see the limit condition we use the formulas $(3.1), (3.2)$
and the corresponding limit condition in the inverse square potential case $(1.1)$
\section{Laplace equation with singular hyperbolic P\"oschl-Teller potential}
\begin{thm} The Dirichlet problem $(1.5)$ for the Laplace equation with singular hyperbolic P\"oschl-Teller potential has the unique solution given by

$$
w(y, x)=\int_0^\infty {\cal P}_\nu^H(y, x , x')w_0(x')d x'\eqno(4.1)
$$
where
$$
  \label{16}
  {\cal P}_\nu^H(y, x, x')= {\cal P}_\nu(Y_2, X_2, x')\eqno(4.2)
$$
with ${\cal P}_\nu$  is the Schwartz integral kernel of the Poisson semi-group $\exp{\left(-Y\sqrt{-L_\nu}\right)}$ with inverse square potential given in $(2.2)$
with  $Y_2$ and $X_2$ as in Lemma 2.2
\end{thm}

{\bf Proof}
We take
 $ V(X_2, Y_2)=\frac{1/4-\nu^2}{X_2^2}$  and set
$$X_2=:\varphi(\xi)+\psi(\eta)=\tan\xi+\tan\eta; Y_2=:\varphi(\xi)-\psi(\eta)=\tan\xi-\tan\eta\eqno(4.3)$$
with $\xi=\frac{x+y}{2}$ and $\eta=\frac{x-y}{2}$ in Lemma 2.2 we obtain
$$\left[\frac{\partial^2 F}{\partial X_2^2}+\frac{\partial^2 F}{\partial Y_2^2}+\frac{1/4-\nu^2}{X_2^2}\right]F=\left((1+\tanh^2\xi)(1+\tanh^2 \eta)\right)^{-1}\left[4\frac{\partial^2 }{\partial \xi\partial\eta}+\frac{1/4-\nu^2}{\sin^2(\xi+\eta)}
\right]F\eqno(4.4)$$
we obtain
$$\left[\frac{\partial^2 }{\partial X_2^2}+\frac{\partial^2 }{\partial Y_2^2}+\frac{1/4-\nu^2}{X_2^2}\right]F=16 \cos^2((x+y)/2)\cos^2((x+y)/2)\left[\frac{\partial^2 }{\partial x^2}+\frac{\partial^2 }{\partial y^2}+\frac{1/4-\nu^2}{16 \sin^2 x}\right]F\eqno(4.5)$$
To see the limit condition we use the formulas $(4.1),(4.2)$
and the corresponding limit condition in the inverse square potential case$(1.1)$
\section{Applications}
In this section we give as an application of our results the following heat Schwartz integral kernels with inverse square and singular trigonometric and hyperbolic P\"oschl-Teller potentials.
That is by using the transmutation formulas of Bragg and Dettman $[3]$ we give explicit solutions to the following heat equation with  inverse square and singular trigonometric and hyperbolic P\"oschl-Teller potentials:\\
{\bf 1)} The heat equation associated to the Schr\"odinger operator with the inverse square potential on $R^+$
 $$  \left \{
    \begin{aligned}
      L_{\nu}U(t, X)=\partial_t U(t, X)=0,
      &\qquad (t, X)\in \mathbb{R_+}\times {\mathbb{R_+}}
      \\
      U(0,X)=U_0(X)\in
      C_0^\infty({\mathbb{R}_+^\ast}).
    \end{aligned}
  \right.\eqno(5.1)$$

{\bf 2) } The heat type equations associated to the Schr\"odinger operators with the singular trigonometric P\"oschl-Teller  potentials
$$
  \left \{
    \begin{aligned}
      L^T_{\nu}V(t, \theta)=\partial_t V(t, \theta),
      &\quad(t, \theta)\in \mathbb{R}_+\times [0,\pi]
      \\
      V(0,\theta)=V_0(\theta)\in
      C_0^\infty([0,\pi]).
    \end{aligned}
  \right.\eqno(5.2) $$

{\bf 3) }The heat type equations associated to the Schr\"odinger operators with the singular hyperbolic P\"oschl-Teller  potentials

$$
  \left \{
    \begin{aligned}
      L^H_{\nu}W(t, x)=\partial_t W(t, x)
      &\quad(t , x)\in \mathbb{R_+}\times {\mathbb{R_+}}
      \\
      W(0,x)=W_0(x)\in
      C_0^\infty({\mathbb{R}_+}).
    \end{aligned}
  \right.
\eqno(5.3) $$
where $L_\nu$, $ L^H_{\nu}$ and $ L^T_{\nu}$ are given in $(1.2)$, $(1.4)$ and $(1.6)$.\\

{\bf Corollary 5.1} The Cauchy problems for the heat equations $(5.1)$, $(5.2)$ and $(5.3)$ has at least formally
 the following Schwartz integral kernels
$$H_\nu(t, X, X') =\frac{1}{4i\sqrt{\pi t}}\int_{\gamma-i\infty}^{\gamma+i\infty}s^{-1/2} e^{st}P_\nu(s^{1/2}, X, X') ds\eqno(5.4) $$
$$H_\nu^S(t, \theta, \theta') =\frac{1}{4i\sqrt{\pi t}}\int_{\gamma-i\infty}^{\gamma+i\infty}s^{-1/2} e^{st}P_\nu^S(s^{1/2}, \theta, \theta') ds\eqno(5.5) $$
$$H_\nu^H(t, X', X') =\frac{1}{4i\sqrt{\pi t}}\int_{\gamma-i\infty}^{\gamma+i\infty}s^{-1/2} e^{st}P_\nu^H(s^{1/2}, x, x') ds\eqno(5.6) $$
where  $P_\nu(Y, X, X'); P_\nu^S(y, \theta, \theta')$ and $ P_\nu^H(y, x, x')$ are the Poisson Schwartz integral kernels
given in $(2.2)$, $(3.2)$  and $(4.2)$ respectively.\\
{\bf Proof} We use essentially the formula $(1.2)$ of $[3]$ and the theorems $(2.1)$, $(3.1)$ and $(4.1)$.

\begin{flushleft}

Universit\'e des Sciences, Technologie et de la M\'edecine(USTM)\\
Facult\'e des Sciences et Techniques\\
D\'epartement de Math\'ematiques et Informatique\\
Unit\'e de Recherche: {\bf \it Analyse, EDP et Modelisation: (AEDPM)}\\
B.P: 5026, Nouakchott-Mauritanie\\
E-mail adresse: khames@ustm.mr\\
\end{flushleft}
\end{document}